\documentclass[]{aa}
\usepackage{graphicx}

\begin{document}

\title{Spectral energy distribution of the $\gamma$-ray microquasar LS~5039}
\authorrunning{Paredes, Bosch-Ramon \& Romero}
\titlerunning{Broad-band emission from LS~5039}
\author{J.~M. Paredes\inst{1} \and V. Bosch-Ramon\inst{1} 
\and G. E. Romero\inst{2,3,}\thanks{Member of CONICET}}

\institute{Departament d'Astronomia i Meteorologia, Universitat de Barcelona, Av. 
Diagonal 647, 08028 Barcelona, Spain; jmparedes@ub.edu, vbosch@am.ub.es.
\and Instituto Argentino de Radioastronom\'{\i}a, C.C.5,
(1894) Villa Elisa, Buenos Aires, Argentina; romero@iar.unlp.edu.ar.
\and Facultad de Ciencias Astron\'omicas y Geof\'{\i}sicas, UNLP, 
Paseo del Bosque, 1900 La Plata, Argentina; romero@fcaglp.unlp.edu.ar.}

\offprints{J.M. Paredes \\ \email{jmparedes@ub.edu}}

\abstract{
The microquasar LS~5039 has recently been detected as a source of very high energy (VHE) $\gamma$-rays.
This detection, that confirms the previously proposed association of LS~5039 with the EGRET source
3EG~J1824$-$1514, makes of LS~5039 a special system with observational data covering nearly all the 
electromagnetic spectrum. In order to reproduce the observed spectrum of LS~5039, from radio to VHE 
$\gamma$-rays, we have applied a cold matter dominated jet model that takes into  account accretion
variability,  the jet magnetic field, particle acceleration, 
adiabatic and radiative losses, microscopic energy conservation in the jet,
and pair creation and absorption due to the external photon fields, as well as the emission from  the
first generation of secondaries.  The radiative processes taken into account are synchrotron, relativistic
Bremsstrahlung and inverse Compton (IC).  
The model is based on a scenario that has been characterized
with recent observational results, concerning the orbital parameters, the orbital variability at X-rays and the
nature of the  compact object. The computed spectral energy distribution (SED) shows a good agreement with
the available observational data.
}

\maketitle

\section{Introduction} \label{intro}

Microquasars are radio emitting X-ray binaries with relativistic radio jets (Mirabel \& Rodr\'{\i}guez~\cite{mirabel99}).
LS~5039 was classified as microquasar when non-thermal radiation produced in a jet was detected with the VLBA (Paredes
et~al.~\cite{paredes00}). New observations  conducted later with the EVN and MERLIN confirmed the presence of an
asymmetric two-sided persistent jet reaching up to $\sim$1000~AU on the longest jet arm (Paredes et~al.~\cite{paredes02},
Rib\'o~\cite{ribo02}). LS~5039 is a high mass X-ray binary (Motch et~al.~\cite{motch97}) whose optical counterpart is a
bright ($V\sim 11$) star of spectral type  O6.5V((f)) (Clark et al.~\cite{clark01}). Although  McSwain et
al.~(\cite{mcswain01}, \cite{mcswain04}) first derived the period and the orbital  parameters of the  system, new values
of the orbital period ($P_{\rm orb}=3.9060 \pm 0.0002$~days, with periastron passage  associated to
$T_0$=HJD~2\,451\,943.09$\pm$0.10), eccentricity ($e=0.35\pm0.04$), phase of periastron passage (0.0), and inclination
(assuming pseudo-synchronization,  $i=24.9\pm2.8^{\circ}$) have been reported by  Casares et al.~(\cite{casares05}), who
place the system at a distance of $2.5\pm0.1$~kpc. The radius of the companion star is moderately close to the Lagrange
point during periastron but without overflowing the Roche lobe. In addition, Casares et al.~(\cite{casares05})  state
that the compact object in LS~5039 is a black hole of mass  $M_{\rm X}=3.7^{+1.3}_{-1.0}$ $M_{\odot}$. At X-rays, the
spectrum of the source follows a power-law and the fluxes are moderate and variable, with temporal scales similar to the
orbital period. The variability is likely associated  with changes in the  accretion rate due to the motion of the
compact object along an eccentric orbit (Bosch-Ramon et~al.~ \cite{bosch05a}).

Among microquasars, LS~5039 is specially relevant for its proposed association with 3EG~J1824$-$1514 (Paredes
et~al.~\cite{paredes00}), an unidentified source in  the 3rd EGRET catalog of high-energy  $\gamma$-ray sources
($>$100~MeV; Hartman et~al.~\cite{hartman99}).  Recently, Aharonian et~al.~(\cite{aha05a}), using the High Energy
Stereoscopic System (HESS), have detected LS~5039 at energies above 250~GeV. This detection, that confirms the
high-energy $\gamma$-ray nature of LS~5039 (Paredes et~al.~\cite{paredes00}), gives strong support to the idea that
microquasars are a distinctive class of high-  and very  high-energy $\gamma$-ray sources. The confirmation of LS~5039 as
a high-energy  $\gamma$-ray source, combined with the fact of being a runaway microquasar ejected from the Galactic plane
at $\sim150$~km~s$^{-1}$ (Rib\'o et~al.~\cite{riboal02}), strongly raises the question about the connection between some
of the remaining unidentified, faint, variable, and soft  $\gamma$-ray EGRET sources and possible microquasar systems
above/below the Galactic plane  (e.g., Kaufman Bernad\'o et al. ~\cite{kaufman02}, Romero et~al.~\cite{romero04},
Bosch-Ramon et~al.~\cite{bosch05b}).

LS 5039 was also proposed to be the counterpart of a COMPTEL source ($\sim$1~MeV) (Strong et~al.~ \cite{strong01}) and
BATSE detected it at soft $\gamma$-rays (Harmon et~al.~\cite{harmon04}). In addition to LS~5039, there is another
microquasar, LS~I~+61~303 (Massi et~al.~\cite{massi04}), that  is likely associated with an EGRET source (Kniffen et
al.~\cite{kniffen97}). Both microquasars have been  extensively studied at different wavelengths, and both have a very
similar spectral energy distribution.  Bosch-Ramon \& Paredes (2004a,b) have explored with a detailed numerical leptonic 
model whether these systems can produce the level of emission detected by EGRET ($>$100~MeV), and the observed
variability. At VHE, Aharonian et al.~(\cite{aha05b})  argue in favor of hadronic origin of TeV photons in case that
$\gamma$-rays are produced within $\sim 10^{12}$ cm,  and Dermer \& B\"ottcher (\cite{dermer05})  point to combined star
IC and synchrotron self-Compton (SSC) emission to explain the different
epoch detections by EGRET and HESS, respectively.

To explain  both, the observed spectrum of LS~5039 from radio to VHE $\gamma$-rays and the variability, we have applied a
broad-band emission model, based on a freely expanding magnetized jet dynamically  dominated by cold protons and
radiatively dominated by relativistic leptons (Bosch-Ramon et~al.~\cite{bosch05c}). In this paper we present the results
obtained after applying such model to LS~5039. The compiled observational data covering the full electromagnetic spectrum
is given in Sect.~\ref{con}. In Sect.~\ref{model} a brief description of the model is provided, and the parameters of the
model are discussed. The results obtained for LS~5039 are presented in Sect.~\ref{disc}, and the work is summarized in
Sect.~\ref{sum}.

\section{Observational data of LS~5039} \label{con}

We compile here the observational data from the literature. The radio data, obtained at 
20, 6, 3.5 and 2 cm wavelengths with the VLA, are from Mart{\'{\i}} 
et~al.~(\cite{marti98}). The optical and infrared data, covering the bands $UBVRIJHKL$, 
are from Clark et al.~(\cite{clark01}) and 
Drilling~(\cite{drilling91}), and have been corrected for absorption ($A_{\rm V}=4.07$, Casares et~al. 
\cite{casares05}). 
The X-ray data (3--30~keV), obtained with RXTE, are from Bosch-Ramon et~al.
~(\cite{bosch05a}), being diffuse background subtracted. 
The (20--430~keV) data, obtained with the BATSE Earth occultation technique, 
are from Harmon et~al.~(\cite{harmon04}). The (1--30~MeV) data are from Zhang~et~al.~(\cite{zhang05}),
obtained with COMPTEL, whose flux values, however, should be taken as an upper limit. This is due to the source 
region defined by the COMPTEL $\gamma$-ray source, GRO~J1823$-$12, at galactic coordinates
($l$=17.5$^{\circ}$, $b=-0.5^{\circ}$), contains in addition to LS~5039/3EG~J1824$-$1514 
two additional EGRET sources (Collmar et~al.~\cite{collmar03}, Zhang et~al.~\cite{zhang05}). 
The high-energy $\gamma$-ray data have been taken from the third EGRET catalog of high-energy 
$\gamma$-ray sources ($>$100~MeV; Hartman et~al.~\cite{hartman99}).
The VHE $\gamma$-ray ($>$100~GeV) data, obtained with HESS, are from 
Aharonian et~al.~(\cite{aha05a}). The values of the magnitudes and fluxes have been transformed in 
luminosities using the updated value for the distance of 2.5~kpc (Casares et al.~\cite{casares05}).
All these observational data are plotted in Fig.~\ref{sed}. 

\subsection{X-ray lightcurve}

The X-ray lightcurve of LS~5039 observed by RXTE, folded in phase with the new orbital period 
value,
shows a clear smooth  peak at phase 0.8 and another marginal one at phase 0.3 (Bosch-Ramon et
al.~\cite{bosch05a}). This lightcurve is hardly explained by accretion through a spherically symmetric
wind  of a typical O type star, 
due to the high velocities of the wind at infinity, and also because of the constraints  on the
stellar mass loss rate ($\la 10^{-6}$~M$_{\odot}$~yr$^{-1}$, Casares et~al.~\cite{casares05}). 
It has been proposed that certain asymmetries in the
wind could produce these peaks in the X-ray lightcurve (Bosch-Ramon et al.~\cite{bosch05a}). 

\section{Modeling} \label{model}

\subsection{An accretion model for LS~5039}

To reproduce approximately the variations in the X-rays, we have assumed the existence of a 
relatively slow equatorial wind  (outflowing velocity $\sim 6\times10^{7}$~cm~s$^{-1}$) 
that would increase the accretion rate 
up to the (moderate) levels required by the observed emission in the context of
our model, and would produce a peak after periastron.
 Although the existence of low velocity flows is not yet well established, it seems that it would not be a rare phenomenon among the massive wind accreting X-ray binaries. This is, for instance, the case of the O9.5V star BD~$+$53$^{\circ}$2790, whose ultraviolet spectrum reveals an abnormally slow wind velocity for its spectral type (Rib\'o et~al.~\cite{riboal05}).
Only a
small fraction of the wind being slower at the equator is enough to increase significantly the accretion  rate
along the orbit if one assumes pseudo-synchronization. 
In this case, 
the orbital plane and the plane perpendicular to companion star rotation axis will be likely the same
(Casares et~al.~\cite{casares05}), allowing for the compact object to accrete from this slow equatorial wind all along the orbit.

\begin{figure}
\centering
\resizebox{\hsize}{!}{\includegraphics{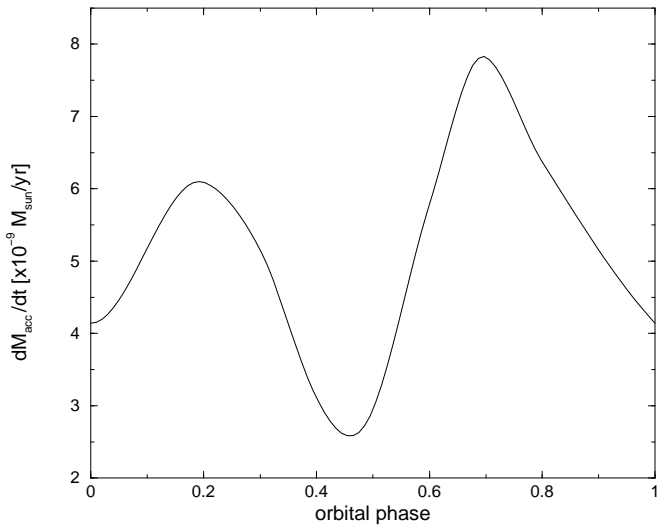}}
\caption{Accretion curve along an orbital period inferred from the observed X-ray lightcurve 
in Bosch-Ramon et~al.~(\cite{bosch05a}), adopting 
the accretion model developed for this work.}
\label{accret}
\end{figure}

To explain the second peak we have
introduced a stream, with a velocity of $5\times10^{6}$~cm~s$^{-1}$ and an additional 
mass loss rate
of $\sim 1$\% of the normal wind, that is formed near periastron passage when the 
attraction force between the two components of the system is at its maximum.
This slower matter stream would produce a very delayed peak at phase 0.7. 
The formation of a matter stream, due to gravitational stress at the smallest 
orbital distances, is not unusual in close binary systems like LS~5039. We
have used the approach followed by Leahy~(\cite{leahy02}) to explain the X-ray lightcurve 
of GX~302--1, also with a peak before periastron passage. 
Although there is still a shift of 0.1 in phase
between the model peaks and the observed ones, it is important to note that
there is likely an accretion disk, and the matter that reaches the disk
spends some time in it until part of this matter is ejected as a
jet  (a lower-limit for the accretion time is the free fall time, implying a shift of 
$\sim$0.06 in phase, when 
falling from 10$^{12}$~cm). 
The accretion rate evolution obtained from the model described here is shown in Fig.~\ref{accret}.

\subsection{A leptonic jet model for LS~5039 broad-band emission}

A new model based on a freely expanding magnetized jet, whose internal energy is dominated by a cold proton plasma
extracted from the accretion disk, has been  developed in a previous work (Bosch-Ramon et~al.~\cite{bosch05c}).
The cold proton plasma and its attached magnetic field, that is frozen to matter and below equipartition, 
provides a framework where internal shocks accelerate a fraction of the leptons up to very high energies. These
accelerated leptons radiate by synchrotron, relativistic Bremsstrahlung with the stellar wind (external) and
within the jet (internal), and IC  processes. In this model, the external seed photons that interact with the
leptons of the jet by IC scattering come from the star, the disk and the corona. A blackbody spectrum is assumed
for the star and the disk, and a power-law plus an exponential cut-off spectrum for the corona. The 
SSC radiation has also been computed as well as  the Bremsstrahlung and Compton self-Compton
radiation, being the radiation of the last two mechanisms negligible.

The dissipated shock kinetic energy that goes to relativistic particles comes from the mean bulk motion kinetic
energy, directly related to the kinetic energy carried by shocks in the plasma. The number of relativistic
particles that can be produced along the jet is constrained by the  limited capability of transferring energy from
the shock itself to the particles, and by the fact that the relativistic pressure must be kept below the cold
proton pressure. The particle maximum energy is limited by the acceleration efficiency and the local energy
losses in the available space. The acceleration efficiency depends on shock properties: shock strength, shock
velocity, magnetic field  direction, and diffusion coefficient (e.g., see Protheroe \cite{protheroe99}). It has
been parametrized due to the lack of knowledge of the specific details of the shock physics for this case. 

\begin{figure}
\centering
\resizebox{\hsize}{!}{\includegraphics{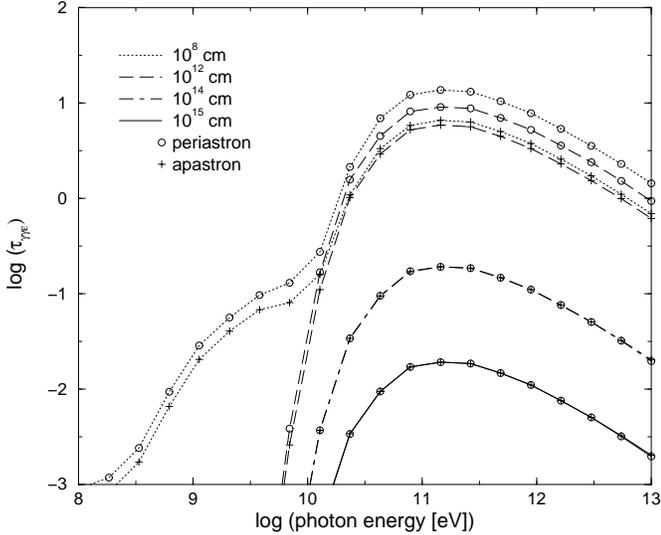}}
\caption{The photon absorption coefficient as a function of the energy 
for photons emitted in the jet 
at different distances from the compact object, at periastron and apastron passages.}
\label{opac}
\end{figure}

In our scenario, the jet formation is mainly constrained  by a characteristic launching radius, where the
ejected matter gets extra kinetic energy from the accretion reservoir, and by the accretion energy 
reservoir itself, which varies along the orbit.  The modeling of the accretion rate changes
allows to introduce variability in a consistent way. Moreover, since the interaction angle between  jet leptons and
star photons changes with the orbital phase, variability naturally appears at very high energies (see
below).  Finally, we have approximately computed the effects produced  on the SED by the photon absorption
under the external photon fields as well as the IC radiation of the first generation of secondaries inside
the binary system. In Fig.~\ref{opac}, we show the photon absorption coefficient as a function of the
photon energy, at different distances from the compact object, during periastron and apastron passages.
It is seen that the corona and disk photon fields (regions of $\sim 10^{8}$ cm) attenuate slightly the
high-energy $\gamma$-ray photons in the inner part of the jet. At VHE, star photon absorption dominates
for distances $\la 10^{13}$~cm, i.e., within the  binary system. At greater distances, absorption goes
down to negligible values. Photon absorption is higher at periastron than at apastron passage, due to the
dependence of the stellar photon density with the orbital distance, although for large distances from 
the  compact object, the opacities at both orbital phases become equal. Our results are similar to
those obtained by B\"ottcher \& Dermer (\cite{bottcher05}) and Dubus (\cite{dubus05}), that present a detailed analysis of the effect
of $\gamma\gamma$ absorption of  VHE $\gamma$-rays near the base of the jet of LS~5039. Moreover,  the
energy dependence of the opacity on the ambient photon fields to the $\gamma$-ray propagation and its
variation with the distance is not much different from that obtained for the microquasar LS~I~+61~303 at
its periastron passage (Romero et al.~\cite{romero05}).

Finally, we note that our model does not take into account the radiative contribution from relativistic protons
that could be accelerated along with the leptons. These protons can cool through interactions with ions from the
stellar wind producing additional $\gamma$-ray emission at high energies (Romero et~al. \cite{romero03},
\cite{romero05}).

\subsection{Parameters of the model} \label{ls}

The model involves a large number of parameters because it aims at reproducing the broad-band spectrum by taking
into account several physical mechanisms. However, in the case of LS~5039 an important fraction of the parameters
are well established from observational data or can be easily estimated from some reasonable assumptions. We call
these parameters `primary parameters'. The remaining `secondary' parameters can be reduced to a manageable
number. In Table~\ref{tab} we present the  primary parameter values of our model.  The first half of them are the
parameters describing the binary system and their components. We adopt updated values taken  from Casares et
al.~(\cite{casares05}). 
The electron power-law index (inferred from  X-ray photon indices found by Bosch-Ramon
et~al.~ \cite{bosch05a}), the jet ejection Lorentz factor (Paredes et~al.~ \cite{paredes02}) and the tangent of
the jet semi-opening angle (Paredes et~al.~\cite{paredes02}) are parameters estimated and/or adopted previously
by other authors. Regarding the jet ejection  Lorentz factor, although it seems well established that the jet is
mildly relativistic (Paredes et~al. \cite{paredes02}), no direct jet velocity estimate has been obtained until
now and we have fixed  the Lorentz factor to 1.1.  The values adopted for the disk inner part temperature and the
corona photon index correspond to  typical values found in other microquasars, because the X-ray spectrum of
LS~5039 only provides clear  upper limits to the luminosities of both disk ($L_{\rm
disk}\simeq10^{34}$~erg~s$^{-1}$) and corona  ($L_{\rm cor}\simeq10^{34}$~erg~s$^{-1}$) (Bosch-Ramon
et~al.~\cite{bosch05a}). The jet-accretion rate parameter, $\kappa$, that gives the ratio between the jet matter
injection rate and the accretion rate, has been taken to be similar to observational and theoretical estimates
(see, e.g., Fender~\cite{fender01}, Hujeirat~\cite{hujeirat04}).

 \begin{table}[]
  \caption[]{Primary parameter values.}
  \label{tab}
  \begin{center}
  \begin{tabular}{ll}
  \hline\noalign{\smallskip}
  \hline\noalign{\smallskip}
 Parameter and unit  &  value \\
  \hline\noalign{\smallskip}
$e$: eccentricity & 0.35$^{a}$ \\
$a$: orbital semi-major axis [cm] & 2.2$\times10^{12}$ $^{a}$\\
$\theta$: jet viewing angle [$^{\circ}$] & 24.9$^{a}$\\
$\dot{m}_{\rm w}$: stellar mass loss rate [$M_{\odot}$~yr$^{-1}$] & $\sim$7$\times10^{-7}$ $^{a}$\\
$M_{\rm x}$: compact object mass [$M_{\odot}$ ] & 3.7$^{a}$\\
$R_{\star}$: stellar radius [$R_{\odot}$] & 9.3$^{a}$\\
$M_{\star}$: stellar mass [$M_{\odot}$]& 22.9$^{a}$\\
$L_{\star}$: stellar bolometric luminosity [erg~s$^{-1}$] &7$\times10^{38}$ $^{a}$\\
$T_{\star}$: stellar surface temperature [K] & $3.9\times10^4$ $^{a}$\\
$p$: electron power-law index & 2.2$^{b}$ \\
$\Gamma_{\rm jet}$: jet injection Lorentz factor & 1.1$^{c}$ \\
$\chi$: jet semi-opening angle tangent & 0.1$^{c}$ \\
$kT_{\rm disk}$: disk inner part temperature [keV]& 0.1 \\
$p_{\rm cor}$: corona photon index & 1.6 \\
$z_0$: Jet initial point [$R_{\rm Sch}$] & 50 \\
$\kappa$: jet-accretion rate parameter & 0.1$^{d}$ \\
  \noalign{\smallskip}\hline
  \end{tabular}
  \end{center}
{
$^{a}$Casares et al.~(\cite{casares05}), $^{b}$Bosch-Ramon et~al.~(\cite{bosch05a}), $^{c}$Paredes et~al.~ (\cite{paredes02}),
$^{d}$Fender~(\cite{fender01}), Hujeirat~(\cite{hujeirat04}). \\
}
\end{table}

 \begin{table}[]
  \caption[]{Secondary parameter values.}
  \label{tab2}
  \begin{center}
  \begin{tabular}{ll}
  \hline\noalign{\smallskip}
  \hline\noalign{\smallskip}
 Parameter and unit  &  value \\
  \hline\noalign{\smallskip}
$\xi$: shock energy dissipation efficiency & 0.5 \\
$f_B$: $B$ to cold matter energy density ratio & 0.01 \\ 
$\eta$: acceleration efficiency & 0.1 \\  
$V_{\rm w}$: equatorial wind velocity [cm~s$^{-1}$] & $6\times10^7$ \\
$\dot{m}_{\rm jet}$: jet injection rate [$M_{\odot}$~yr$^{-1}$]&
 $4\times10^{-10}$\\
$\zeta$: max. ratio hot to cold leptons & 0.1 \\ 
$r_{\rm l}$: launching radius [$R_{\rm Sch}$] & 9 \\
$V_{\rm stream}$: periastron stream velocity [cm~s$^{-1}$] & $5\times10^6$ \\

  \noalign{\smallskip}\hline
  \end{tabular}
  \end{center}
\end{table}

Next, we describe the secondary parameters of the model, which have been adopted in order to better
reproduce different spectral and variability properties of the source.
They are summarized
in Table~\ref{tab2}. The shock energy dissipation efficiency, $\xi$, that gives the energy dissipated by
shocks in the jet, has been taken (Bosch-Ramon et~al.~\cite{bosch05c}) to be at most the half of the kinetic energy of the
shock to get enough luminosity at energies beyond X-rays\footnote{This does not mean that
half of the shock kinetic energy is dissipated through shocks, but that it might 
be so in case that the energy losses would require it, i.e., the shock could be radiatively
efficient up to a 50\% of the kinetic energy in the regions where energy losses
are very strong (the total jet energy lost through radiation is about 15\%).}. 
The ratio of the magnetic field ($B$) energy density to the cold matter
energy density has been fixed to $f_B=0.01$ taking into account X-ray flux constraints from observations
(Bosch-Ramon et~al. \cite{bosch05a}). The acceleration efficiency parameter, $\eta$ ($\times q_{\rm
e}Bc$), has been fixed such that the maximum energy photons reach the energies at which the source has
been detected by HESS. This value for $\eta$ is in agreement with that of a strong and
transrelativistic shock with a magnetic field parallel to its direction of motion and a diffusion
coefficient close to the Bohm limit (Protheroe~\cite{protheroe99}). The stellar wind velocity 
in the vicinity of the compact object ($V_{\rm w}$), estimated in Sect. \ref{con}, is likely affected by peculiarities of
the star (e.g., close
to but not suffering Roche-lobe overflow, fast rotation, etc; see Casares et~al.~ \cite{casares05}).
This velocity implies, along with the known stellar mass loss rate (see Table~\ref{tab}) and
assuming isotropic flux, a mean accretion rate along the orbit of $\dot{m}_{\rm
acc}\sim4\times10^{-9}$~$M_{\odot}$~yr$^{-1}$ (i.e. $\sim 5$\% of the Eddington rate). 
From  this value  and $\kappa$, the jet matter injection
rate can be derived: $\dot{m}_{\rm jet}\sim4\times10^{-10}$~$M_{\odot}$~yr$^{-1}$. 
In order to try to satisfy both, the assumption that the jet is 
cold matter
dominated and the observed optically thin nature of the radio spectrum, we have fixed 
the maximum number of relativistic
leptons\footnote{The number of relativistic leptons is fixed by $\xi$, $\zeta$, and the condition 
that their relativistic pressure cannot dominate the cold particle pressure.} 
to be $\zeta=0.1$ of the total number of particles in the jet. This is higher
 than the value used by
Bosch-Ramon et~al.~(\cite{bosch05c}). The implications of this are discussed in Sect.~\ref{disc}.
Finally, the launching radius can be estimated to be $\sim 9 R_{\rm
Sch}$, since  the energy to eject the jet is assumed to be accretion energy transfered and the Lorentz
factor is roughly known (Bosch-Ramon et~al.~\cite{bosch05c}).

 We note that we are not fitting data but reproducing a broadband spectrum as a whole. This implies that the values of the parameters used by the model are approximated.
The parameters in Tables~\ref{tab} and \ref{tab2} not obtained directly from observations have a range of validity within 10\% for $p$, $\Gamma_{\rm jet}$, $kT_{\rm disk}$, $p_{\rm cor}$, $V_{\rm w}$ and $V_{\rm
stream}$; and within a factor of a few for $z_0$, $\kappa$, $\xi$, $f_B$, $\eta$,  $\dot{m}_{\rm jet}$, $\zeta$
and $r_{\rm l}$.

\section{Results} \label{disc}

\subsection{Spectral energy distribution of LS~5039}

\begin{figure*}
\centering
\resizebox{\hsize}{!}{\includegraphics{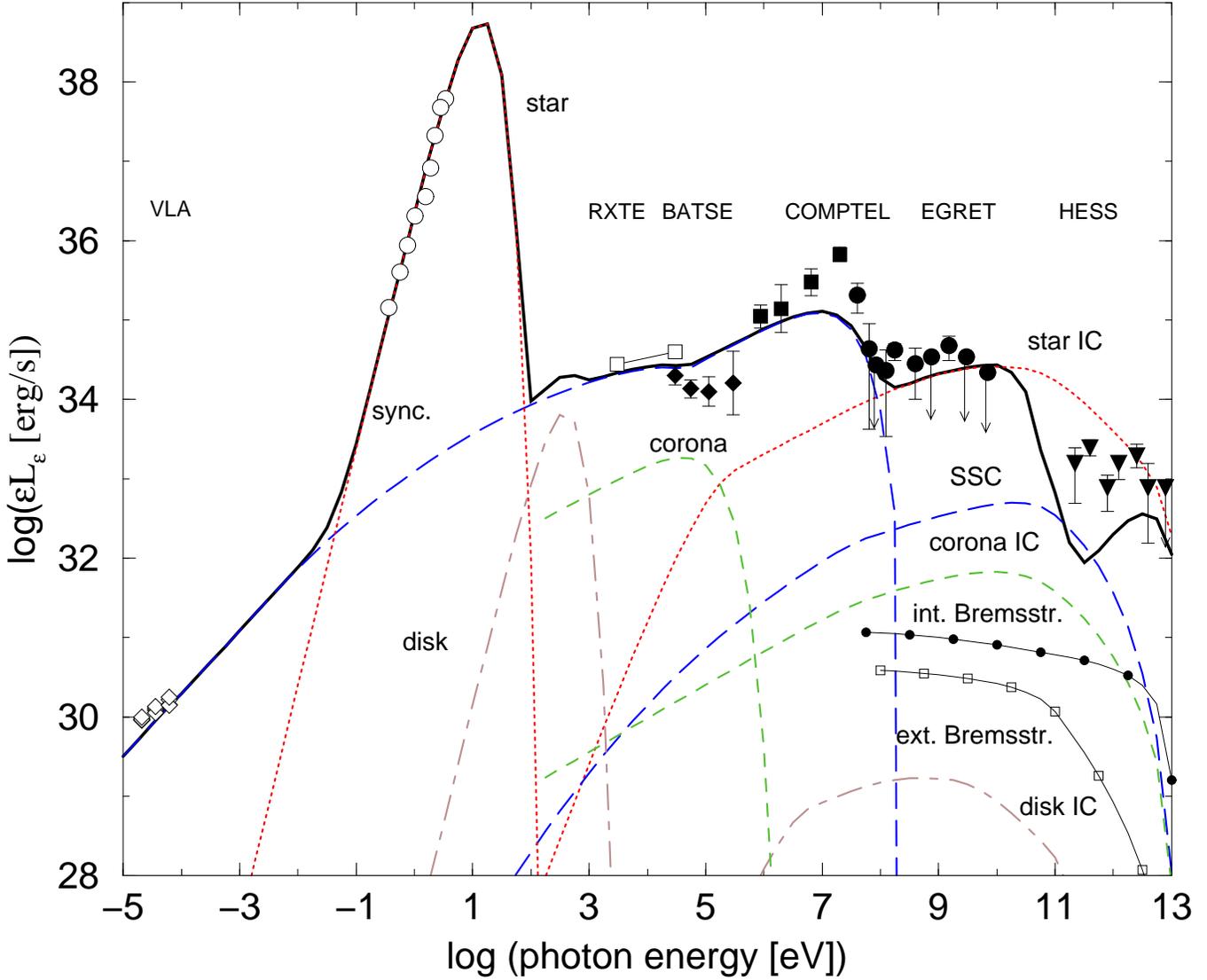}}
\caption{
Comparison of the spectral energy distribution of LS~5039 computed 
from the present model, at the periastron passage and 
using the parameters of Table~\ref{tab}, with the observed data. We show the different 
components of the emission as well as the sum of all of the them, which is attenuated 
due to $\gamma\gamma$ absorption (solid line). The points observed are 
from Mart{\'{\i}} 
et~al.~(\cite{marti98}) (VLA, diamond), Clark et al.~(\cite{clark01}) and 
Drilling~(\cite{drilling91}) (optical, circle, corrected of absorption), Bosch-Ramon et~al.
~(\cite{bosch05a}) (RXTE, square), 
Harmon et~al.~(\cite{harmon04}) (BATSE, diamond filled), Collmar et~al.~(\cite{collmar03}) 
(COMPTEL, square filled), 
Hartman et~al.~(\cite{hartman99}) (EGRET, circle filled) and  Aharonian et~al.~(\cite{aha05a})
(HESS, triangle down filled). The arrows in the 
EGRET and HESS data represent upper limits (3$\sigma$). 
}
\label{sed}
\end{figure*}

Figure~\ref{sed} shows the computed SED (continuous thick line) of LS~5039 at the periastron 
passage (phase 0.0) using the physical parameters listed in Tables~\ref{tab} and \ref{tab2}. 
We have plotted
also the contributions from the star, disk, corona and synchrotron radiation as well as the
inverse Compton emission  and relativistic Bremsstrahlung. In the same 
figure we plot also the observational data. 

There is a general agreement between the model and the observed data, although there are some
discrepancies that require some explanation. First at all, we note that the multiwavelength data were not
taken simultaneously, and this can be an important source of discrepancy for any time-dependent model.

 At the radio band, 
the computed spectrum has a spectral index of 0.3, instead of the observed $\sim$0.5 (Mart{\'{\i}} et~al.~\cite{marti98}; 
where $F_{\nu}\propto\nu^{-0.5}$). It seems that an
additional process, not contemplated in our scenario, rises up the number  of electrons emitting optically thin
radio emission (e.g., interactions with the wind or the environment). Concerning the spatial distribution of radio
emission along the jet, our computed fluxes at 5~GHz are about 10, 6 and 0.1~mJy for the inner part of the jet ($<$ 1~AU), at
middle distances from the origin (1--1000~AU) and beyond ($>$ 1000~AU), respectively.
This roughly reproduces the
extended emission of the radio source observed using VLBI techniques (Paredes et~al.~\cite{paredes02}). 

At the X-ray band, most of the emission is synchrotron radiation coming from
the relativistic electrons of the jet and only a marginal contribution comes
from the disk and almost a null contribution from the corona components  (see
also, e.g., Markoff et~al. \cite{markoff01}). 
 Although the XMM-Newton data does not show evidences of a thermal disk component (Martocchia et~al. \cite{martocchia05}), we have adopted in our model a disk with moderate luminosity. The role played in the model by this component is negligible, and its contribution appears in Fig.~\ref{sed} as a small bump over the power-law spectrum below 1~keV.
The computed X-ray luminosities 
and photon indices are very close to those detected so far (see
next section and also Bosch-Ramon et~al.~\cite{bosch05a}).

Gallo et~al. (\cite{gallo03}) have proposed that radio emitting X-ray binaries,
when in the low-hard state, follow a particular correlation concerning radio
and X-ray luminosities. 
 In the context of this correlation, LS~5039 is pretty under-luminous 
at X-rays, presenting a permanent low-hard-like X-ray state.
This could be explained by the fact that its radio emission is
optically thin, coming mainly from regions outside the binary system, instead
of being optically thick and completely core dominated (as in the case of models such as
the one presented by Markoff et~al. \cite{markoff01}, see also Heinz \& Sunyaev~\cite{heinz03}). 
As noted above, it might be qualitatively explained by the increase in relativistic electrons
predicted by our model, although the issue requires further 
investigation since our radio  spectrum
is still too hard.

At BATSE and COMPTEL energies, the main contribution still comes from synchrotron emission of the jet.
There is a small disagreement between the predicted luminosity and 
the one observed by BATSE. However, we note that these data are averaged along many days
(Harmon et~al.~\cite{harmon04}), whereas the plotted SED is computed at
phase 0.0, when the accretion is significant though not the highest.
At the COMPTEL energy range, as noted above, the moderate disagreement between the
computed and observed data is likely produced by the fact that the MeV radiation
coming from the
COMPTEL source GRO~J1823$-$12, to which LS~5039 is associated, might be a
superposition of the MeV emissions of the three different EGRET sources that
are in the field  of view (Collmar et~al.~\cite{collmar03}).  This means that
the data from COMPTEL should be actually taken as an upper limit. In such a case, our
computed values would be consistent with the data.

At high-energy $\gamma$-rays (0.1--10~GeV), the computed luminosities are similar to those observed by
EGRET while the computed photon indices are slightly harder. This fact could be naturally associated
with the low temporal resolution of EGRET, with the shortest viewing periods spanning for one week and
providing time averaged spectra that could mask harder ones. We note however that, because of the
possible effect of pair creation in the jet, the slopes of the SEDs at high-energy  $\gamma$-rays,
unlike the computed luminosities, are not very reliable (see Bosch-Ramon et~al.~ \cite{bosch05c}).
The dominant component at this band of the spectrum is  star IC, whereas the synchrotron radiation is only
important at the lower part of EGRET energies. 
The emission of secondary pairs created within the binary system is not significant.

 At VHE $\gamma$-rays, star IC emission dominates. The spectrum reaches several TeV, and it is limited by the  maximum
particle energy, which can reach $\sim 10$~TeV.  The contribution of the star IC scattering  is favored during the
periastron passage due to a higher photon density as well as  a smaller interaction angle between the electrons and the
photons, which influences the IC scattering.  The contribution from  the
the SSC component is minor, although is more significant at apastron   
than at periastron.
The VHE spectrum is strongly affected by the $\gamma\gamma$ opacity  due to pair creation
within the binary  system, being the absorption stronger at the periastron passage. It appears therefore unavoidable an
important decrease of the emission as well as an additional source of  variability at those energies. Also, a hardening of
the emission, as seen in Fig.~\ref{sed}, could take place. Our model underestimates the TeV emission, although the
difference is less than one order of magnitude. Possibly, some high energy processes involving particle acceleration and
emission are taking place in the jet at middle scales, where the 100 GeV photon absorption is not important. This could
also be related to the optically thin nature of the radio emission. This outer emission would not be as variable as the
emission from the compact jet.  In any case, a deeper treatment at this energy band concerning VHE emission and pair
creation phenomena is required  (Khangulyan \& Aharonian~\cite{khangulyan05}).

The results obtained by applying other leptonic models (e.g. Bosch-Ramon et~al. \cite{bosch05b}, Dermer \& B\"ottcher \cite{dermer05}) do not differ substantially from the comprehensive model presented here, which can explain the broad-band emission from LS~5039. Otherwise, hadronic models can reproduce the emission at VHE energies (Romero et~al. \cite{romero03}) though it is hard to explain how typical EGRET fluxes could be achieved (Romero et~al. \cite{romero05}). Comparing both types of model, it seems that the leptonic ones can explain better than the hadronic ones the broadband spectrum at least up to $\sim 1$~GeV, being both suitable at higher energies since the energetic requirements are not as strong as those to explain the whole emission from the source.

\subsection{Variability properties of LS~5039} \label{resls}

\begin{figure}
\centering
\resizebox{\hsize}{!}{\includegraphics{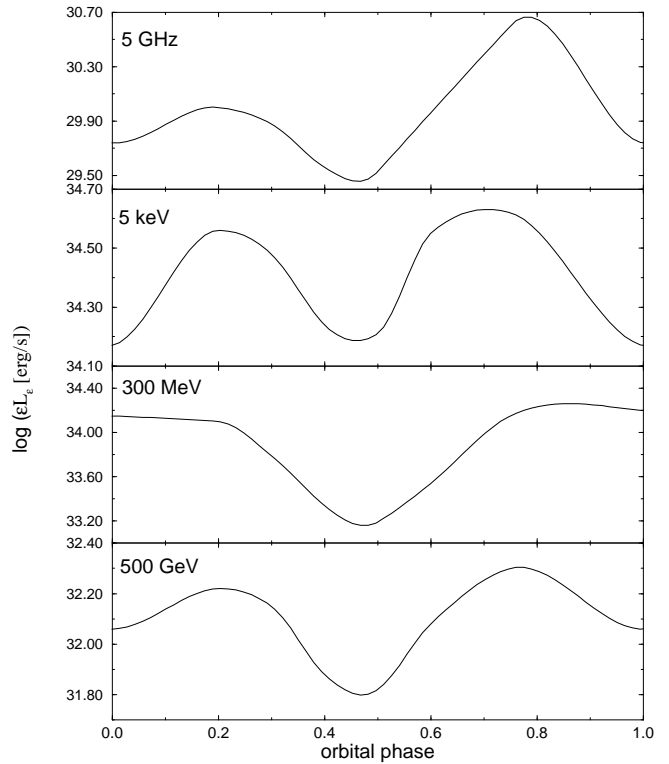}}
\caption{
The panel shows the flux variations 
along the orbit for the four energy bands considered here. We note that 
emission 
at very high-energy $\gamma$-rays is strongly attenuated by the photon absorption 
in the star photon field.
}
\label{var}
\end{figure}

Once the accretion model has been applied to reproduce roughly the X-ray lightcurve, we explore what are the consequences
at the other bands of the spectrum. In Fig.~\ref{var} we  show the flux  variations along the orbit at four energy
bands.  At radio and X-ray energies, the predicted flux variations are well correlated with accretion, since they have
similar shapes to the accretion orbital curve (see Fig.~\ref{accret}). The predicted amplitude variations are of about
one order of magnitude in radio and  a factor of three at X-rays, although we note that, if the radio emitting region is
large enough, say $10^{15}$~cm, the variations at these wavelengths will be significantly smoothened by light crossing
time limitations (as it seems to be the case, see Rib\'o~(\cite{ribo02}). 
The variations of the radio spectral indices along the orbit are not treated here, due to the model limitations at radio. 
However, we mention that the X-ray photon index
shows a trend similar to that found by Bosch-Ramon et~al.~(\cite{bosch05a}), i.e. higher fluxes when harder spectra.
Nevertheless, the anticorrelation is not as strong as that observed.
Emission at EGRET energies varies almost
one order of magnitude, although the shape of the lightcurve is different because of the higher rate of IC interaction during phase 0.0 due to
angular effects. At VHE, however, the emission is strongly absorbed by the stellar field, producing a dip around
periastron passage. At phases 0.6--0.9, there is a long standing peak that is related to the stream related peak at phase
0.7 (0.8) that is not close enough to the stellar companion for banishing due to $\gamma\gamma$ absorption. A similar
peak seems to be present in HESS data (Aharonian et~al.~\cite{aha05a}) when folded in phase with the new ephemeris of
Casares et~al.~(\cite{casares05}), although further observations should be carried out to confirm this. At phases
0.2-0.3, a smaller peak is seen, which might be detected eventually in future VHE observations. We note that, as it is
seen in Fig.~\ref{opac}, if TeV emission comes from distances typically larger than the semi-major orbital  axis it would
not be significantly attenuated by photon absorption. In such a case, variability would be limited by the size of the
emitting region.

\section{Comments and conclusions} \label{sum}

We have applied to the microquasar LS~5039 a cold matter dominated jet model that takes into 
account accretion variability,  the jet
magnetic field, particle acceleration, microscopic energy conservation in the jet, and  pair 
creation and absorption due to the external and internal photons, as well as the emission from 
the first generation of secondaries under the external photon fields.

The computed SED from radio to VHE $\gamma$-rays 
has been compared with the available observational data, gathered at different bands of
the electromagnetic spectrum, showing in general a good agreement. Although the model is limited by its 
phenomenological 
nature, we can extract some general conclusions that could be useful to understand the physics 
behind these objects as well as for going deeper in more basic theoretical aspects concerning the jet 
plasma characteristics, particle acceleration and jet formation phenomena.
In particular, in the context of our model, we can state 
that the jets are radiatively efficient ($\sim 15$\%) 
and sites of particle acceleration up to multi-TeV energies. Otherwise, accretion processes are radiatively inefficient.

The strong absorption of VHE $\gamma$-rays by the stellar photon field seems to make hard the detection of high fluxes of
radiation if they are generated in the inner regions of the  jet, closest to the compact object. Possibly, since the
source is not very faint at these energies, some high energy processes involving particle acceleration and emission are
taking place in the jet at  middle scales, where the 100 GeV photon absorption is not important. This could also be
related to the high fluxes and optically thin nature of the radio emission. Variability at these energy bands could
provide information about the location of the particle acceleration processes and emitting regions.

Changes in the accretion rate due to the orbital eccentricity and a particular wind 
density profile can explain the spectral and variability properties observed at X-rays.
It hints to a companion stellar wind that is not spherically symmetric nor fast
in the equator, since otherwise the $\gamma$-ray radiation could not be powered.

We conclude that, with reasonable values for the different parameters, our model seems to be good enough as to
describe the general features of the spectrum of LS~5039 from radio to TeV $\gamma$-ray energies, the leptonic
model being a reasonable candidate for explaining the $\gamma$-ray emission on energetic grounds.

\begin{acknowledgements}
We thank Mark Rib\'o for fruitful discussions as well as useful comments and 
suggestions on the subject studied here.
J.M.P. and V.B-R. acknowledge partial support by DGI of the
Ministerio de Educaci\'on y Ciencia  (Spain) under grant AYA-2004-07171-C02-01, as well as additional
support from the European Regional Development Fund (ERDF/FEDER). During this work, V.B-R has been
supported by the DGI of the Ministerio de (Spain) under the fellowship BES-2002-2699. G.E.R is
supported by the Argentine Agencies CONICET (PIP 5375) and ANPCyT (PICT 03-13291).  
\end{acknowledgements}

{}

\end{document}